\documentclass[preprint,onecolumn,nofootinbib]{revtex4}

\usepackage{mathrsfs}  
\usepackage{amssymb}
\usepackage{slashed}  
\usepackage[graphicx]{realboxes}  
\usepackage{tikz}  
\usepackage{subcaption}  
\usepackage{pgfplots}  
\usepackage{booktabs}  
\usepackage{comment} 

\usepackage[colorlinks=true,linkcolor=blue,urlcolor=blue,filecolor=black,citecolor=red,pdfstartview=FitV,pdftitle={},pdfsubject={},pdfkeywords={},pdfpagemode=None,bookmarksopen=true]{hyperref}  
\usepackage{graphicx}
\usepackage{amsmath}
\usepackage{amsfonts}  
\usepackage{amssymb,ulem}
\usepackage{color,xcolor}%
\usepackage{CJK}
\usepackage{amsthm,amsmath,amssymb} 
\usepackage{mathrsfs}
\usepackage{url}
\usepackage{hyperref}
\usepackage{array}

\usepackage{dcolumn}
\usepackage{float}
\usepackage{appendix}
\begin{document}

\captionsetup{justification=raggedright}

\title{Quasinormal modes of a $d$-dimensional regular black hole featuring an integrable singularity}

\author{Zhongzhinan Dong}
\thanks{dzzn\_edu@163.com}
\author{Dan Zhang}
\thanks{danzhanglnk@163.com}
\author{Guoyang Fu}
\thanks{FuguoyangEDU@163.com}
\author{Jian-Pin Wu}
\thanks{jianpinwu@yzu.edu.cn} 
\affiliation{
 Center for Gravitation and Cosmology, College of Physical Science and Technology, Yangzhou University, Yangzhou 225009, China
}
	
\begin{abstract}
		
In this paper, we exhaustively investigate the quasinormal modes (QNMs) of a probe scalar field over a $d$-dimensional regular black hole (BH) characterized by the parameter $A$. The quasinormal frequencies (QNFs) exhibit different behaviors with respect to the parameter $A$ for $d=4$ and $d>4$. Firstly, the trends of QNFs with respect to $A$ exhibit completely opposite patterns for the case of $d=4$ and $d>4$. Secondly, in the $4$-dimensional regular BH, a non-monotonic behavior with respect to $A$ is observed in the imaginary part of the fundamental modes with vanishing angular quantum number. In contrast, for nonzero angular quantum number or $d>4$, non-monotonic behavior is observed only in the overtones. Thirdly, an overtone outburst accompanied by an oscillatory patter is observed only in the case of $d>4$, but not in $d=4$.

\end{abstract}
	
\maketitle
\tableofcontents

\section{Introduction}

Regular black holes (BHs) were initially proposed as a solution to avoid the core singularity that is inherent in ordinary BHs. The scalar curvature of these regular BHs remains finite everywhere, including at the core. In contrast to ordinary BHs, regular BHs have complete null and timelike geodesics in their spacetime. These BHs can be formed by incorporating quantum gravity effects \cite{Bardeen:1968,Hayward:2005gi,Frolov:2016pav,Ling:2021olm}, exotic matter like the nonlinear Maxewell field \cite{Ayon-Beato:1998hmi,Dymnikova:2004zc,Cataldo:2000ns} that typically defy the standard energy conditions. For comprehensive reviews on regular BHs, refer to  \cite{Bambi:2023try,Vagnozzi:2022moj,Lan:2023cvz}.

Regular BHs can be usually classified into two categories based on their asymptotic behavior near the center: those with a de-Sitter (dS) core and those with a Minkowski core. Regular BHs with a dS core feature an energy density that converges to a non-zero constant as one approaches the center of the BH, specifically, $\displaystyle \lim_{r\rightarrow0}\rho(r)=\rho_0 \neq 0$. These consist of the widely recognized Bardeen BH \cite{Bardeen:1968}, Hayward BH \cite{Hayward:2005gi} and Frolov BH \cite{Frolov:2016pav}. We would like to mention that in \cite{Boos:2023icv}, the authors introduce a novel regular BH solution, which greatly extends the Hayward class. This solution incorporates mass-dependent regulators, resulting in significant, percent-level effects on observables for regular astrophysical BHs. 

The BH with Minkowski core was first proposed in \cite{Xiang:2013sza} by incorporating an exponentially suppressed Newton potential. This solution has been further extended to encompass a category of regular BHs featuring various kinds of exponential potentials \cite{Culetu:2013fsa,Culetu:2014lca,Rodrigues:2015ayd,Simpson:2019mud,Ghosh:2014pba,Ghosh:2018bxg,Li:2016yfd,Martinis:2010zk,Ling:2021olm}. In contrast to regular BHs with a dS core, those with a Minkowski core have an energy density that approaches zero at the center, specially, $\displaystyle \lim_{r\rightarrow0}\rho(r)=0$. 

These regular BHs typically possess distinct outer and inner horizons, each exhibiting nonzero surface gravity. The presence of a nonzero inner horizon surface gravity typically leads to an exponential growth of the gravitational energy within the vicinity of the inner horizon. This phenomenon is known as the mass inflation instability \cite{Poisson:1990eh,Brown:2011tv}. In order to circumvent the mass inflation instability, the authors in \cite{Carballo-Rubio:2022kad} propose a regular BH model where a surface gravity becomes zero at the inner horizon while preserving non-zero at the outer horizon. In contrast to scenarios with finite surface gravity at the inner horizon, this model does not exhibit an exponential growth associated with mass inflation instability \cite{Carballo-Rubio:2022kad}.

Recently, an alternative approach to avoid the problem of mass inflation instability was proposed in \cite{Casadio:2023iqt,Ovalle:2024wtv}. The key idea is to incorporate quantum matter described by regular wave functions, resulting in a novel regular BH geometry with an integrable singularity\footnote{Integrable singularities are defined as the locations where the effective energy-momentum tensor and curvature invariants become infinite, but their ``volume'' integrals are still finite \cite{Lukash:2013ts}.} and no Cauchy horizon in spherically symmetric configurations. Following Ref. \cite{Casadio:2023iqt}, we refer to this novel regular BH as the one with localized sources of matter (LSM). In contrast to those regular BHs that possess clearly defined outer and inner Cauchy horizons, this novel regular BH proposed in \cite{Casadio:2023iqt} allows radial geodesics to reach the core of the BH. Subsequently, this solution was extended to encompass higher-dimensional scenarios \cite{Estrada:2023dcj}. In this paper, we will study the characteristics of the quasinormal modes (QNMs) of a probe scalar field over this novel d-dimension regular BH.

During the ringdown phase, the BH emits gravitational waves (GWs) with characteristic discrete frequencies known as quasinormal frequencies (QNFs). The frequencies encode information regarding the decaying scales and damped oscillations of the BH \cite{Berti:2009kk}. Studying the properties of QNFs offers an opportunity to implement the strong field tests of general relativity (GR) \cite{LIGOScientific:2021sio,Berti:2015itd,Berti:2018vdi}. Additionally, it allows for testing the nature of compact remnants formed after a coalescence \cite{Cardoso:2019rvt} as well as potential imprint of quantum gravity \cite{Fu:2022cul,Fu:2023drp,Gong:2023ghh,Zhang:2024nny,Song:2024kkx}. Nevertheless, most regular BHs are commonly formed by integrating quantum gravity effects at the phenomenological level, which complicates the task of establishing consistently effective equations for gravitational perturbations. Fortunately, even when merely a probe matter field is taken into account over these regular BHs, their QNM spectra are likewise affected by the background spacetime. As a result, these QNM spectra may be utilized to phenomenologically simulate the GW form during the ringdown phase and also offer vital insights into the internal structure of the BH \cite{Berti:2005ys,Berti:2018vdi,Fu:2022cul,Fu:2023drp,Gong:2023ghh,Moura:2021eln,Moura:2021nuh,Moura:2022gqm,Lin:2024ubg,Ghosh:2022gka,Zhang:2024nny,Song:2024kkx,Zhu:2024wic,Konoplya:2024lch,Konoplya:2024hfg,Gingrich:2024tuf}.

At first, the majority of research efforts concentrated on the fundamental modes of QNM spectra. Nevertheless, the examination of GW data from LIGO/Virgo has lately shown compelling evidence in favor of the presence of overtone patterns \cite{Isi:2020gwy,LIGOScientific:2020tif}. Compared to results obtained exclusively from the fundamental modes, incorporating overtones theoretically provide more exact information regarding the mass and spin of the remnant BH \cite{Giesler:2019uxc}. Moreover, several studies have emphasized the importance of overtones in accurately simulating the ringdown phase \cite{Giesler:2019uxc,Oshita:2021iyn,Forteza:2021wfq,Oshita:2022pkc}. These results challenge the prevailing belief that the fundamental mode dominates the signal primarily and imply that the quasinormal ringing might begin earlier than previously anticipated.

Furthermore, a recent study has revealed a connection between the properties of the high overtones and the geometric structure surrounding the event horizon of a BH \cite{Konoplya:2022pbc}. The research discovered that even a slight modification near the event horizon might have a substantial influence on the first few overtones \cite{Konoplya:2022pbc}, so offering an opportunity to explore the structure of the event horizon. In addition, in constrast to echoes, overtones display a significantly higher energy contribution, rendering them an essential domain for investigating diverse BHs \cite{Konoplya:2022hll,Konoplya:2022iyn,Konoplya:2023aph}. These significant discoveries offer additional impetus for our ongoing exploration of the overtones of different BHs. Considering the aforementioned motivations, we will study the characteristics of both the fundamental modes and the high overtone modes of the scalar perturbation of the novel regular BHs \cite{Estrada:2023dcj}. In addition to studying the implications of the model's characteristic parameters, we will also investigate the influence of extra dimensions.

This paper is structured as follows. In Section \ref{section2}, we provide a brief review of this regular BH with LSM, derive the expression for the scalar field equation over this $d$-dimensional BH, and then discuss the properties of the corresponding effective potentials. In Section \ref{section3}, we use the pseudo-spectral method (PSM) to obtain the QNFs and analyze their characteristics. The conclusions and discussions are presented in Section \ref{section4}. We also provide a detailed introduction to the PSM in Appendix \ref{app:A}. 

\section{Scalar field over the regular black hole}\label{section2}

Following the framework presented in \cite{Casadio:2023iqt}, the authors of \cite{Estrada:2023dcj} construct a $d$-dimensional analytical regular BH featuring an integrable singularity and no an internal horizon by incoporating the LSM.
The model's exterior geometry can be written as \cite{Estrada:2023dcj}:
\begin{eqnarray}
&&
ds^2=-f(r)dt^2+f(r)^{-1}dr^2+r^2d\Omega_{d-2}\,,
 \nonumber
	\\
	&&
f(r)=1-\frac{2G_d m(r)}{r^{d-3}}\,,~~~~\;
\label{metric}
\end{eqnarray}
where $m(r)$ is the mass function with
\begin{eqnarray}
m(r)=\frac{Mr^{d-3}}{A+r^{d-3}}\,. \label{mr}
\end{eqnarray}
Here, $ d\Omega_{d-2} $ represents the transversal section of a $(d-2)$-dimensional sphere, $ G_d $ denotes the $d$-dimensional Newton constant, $A$ is a positive constant measured in units of $\ell_p^{d-3} $, and $M$, quantified in units of $\ell_p^{-1}$, corresponds to the mass parameter, where $\ell_p$ refers to Planck units. Then, the energy density can be explicitly calculated as \cite{Estrada:2023dcj}:
\begin{eqnarray}
\rho=\frac{d-2}{8\pi}\frac{A(d-3)M}{r^2(A+r^{d-3})^2}\,. \label{rho}
\end{eqnarray}
From the above equations \eqref{mr} and \eqref{rho}, it is evident that the mass function $m$ and the energy density $\rho$ exhibit behaviors of $m\sim r^{d-3}$ and $\rho\sim r^{-2}$ near the origin, respectively. The asymptotic behaviors of $m$ and $\rho$ guarantee that there is no internal horizon linked to an unstable core, and that the energy-momentum tensor and Ricci scalar are integrable at the origin. In addition, it can be shown that the radial timelike geodesic and the tidal forces of this model are finite \cite{Estrada:2023dcj}. Furthermore, the model's central integrable singularity is weak. This gravitationally weak singularity hence will not crush any object following a radial timelike world-line.

\begin{figure}[htbp]
	\centering
	\includegraphics[width=0.6\textwidth]{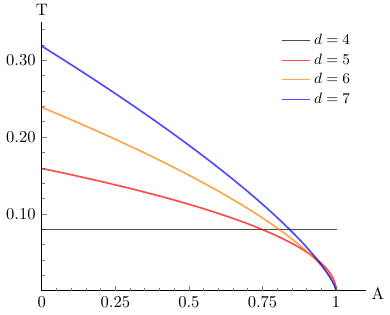 }
	\caption{The Hawking temperature $T$ as a function of the parameter $A$ in different dimensions.}
	\label{fig:9}
\end{figure}

The Hawking temperature of this regular BH can be calculated as follows:
\begin{eqnarray} 
T=\frac{(d-3)M}{2\pi}\frac{r_h^{d-4}}{(A+r_h^{d-3})^2}\,.\label{temperature} 
\end{eqnarray} 
Here, $r_h=(2M-A)^{1/(d-3)}$, with $A\leq 2M$, is the event horizon of the BH and is determined by taking the positive root of $f(r)=0$. In this paper, without loss of generality, we set $M=1/2$, implying $A\leq 1$. From Eq. \eqref{temperature}, it is evident that when $d=4$, the Hawking temperature approaches a constant value as $A$ tends to $1$. Nevertheless, for dimensions greater than $4$ ($d>4$), the Hawking temperature decreases towards zero as $A$ approaches $1$. To visualize this picture, we present the Hawking temperature as a function of the parameter $A$ in Fig.\ref{fig:9}.

\begin{figure}[htbp]
	\centering
	\begin{minipage}[b]{0.45\textwidth}
		\centering
		\includegraphics[width=\textwidth]{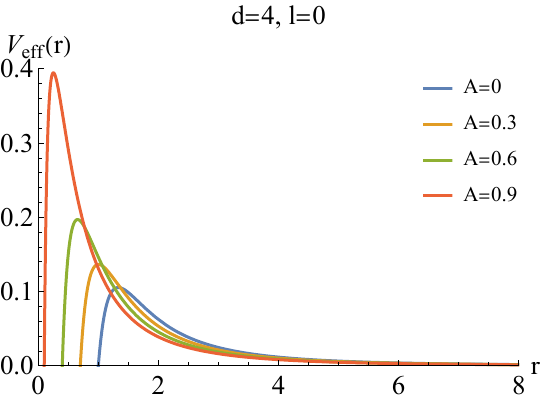 }
	\end{minipage}
	\hspace{1em} 
	\begin{minipage}[b]{0.45\textwidth}
		\centering
		\includegraphics[width=\textwidth]{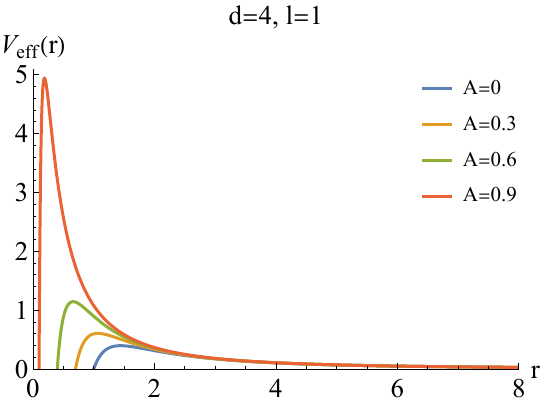 }			
	\end{minipage}
	\caption{The effective potential $V_{eff}(r)$ for $ d=4 $ with different parameter $A$}
	\label{fig:7}
	\vspace{10pt} 
\end{figure}

\begin{figure}[htbp]
	\centering
	\begin{minipage}[b]{0.45\textwidth}
		\centering
		\includegraphics[width=\textwidth]{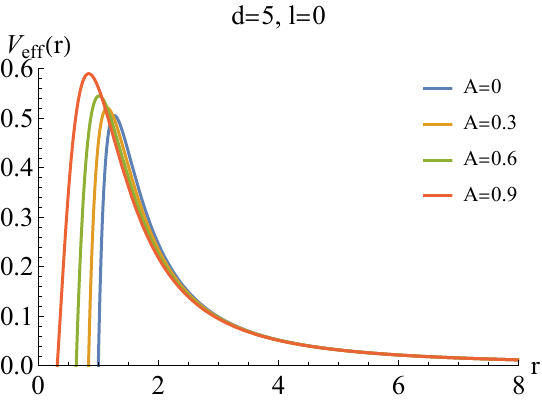}
	\end{minipage}
	\hspace{1em} 
	\begin{minipage}[b]{0.45\textwidth}
		\centering
		\includegraphics[width=\textwidth]{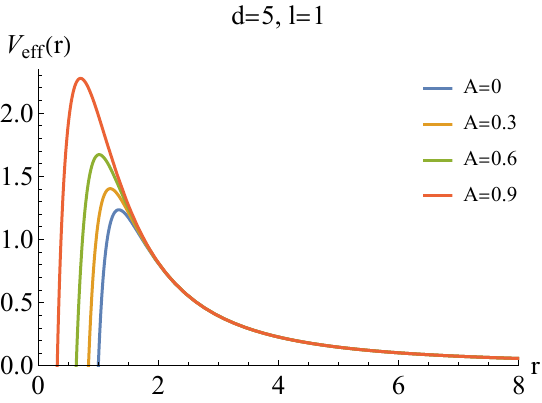}			
	\end{minipage}
	\caption{The effective potential $V_{eff}(r)$ for $ d=5 $ with different parameter $ A $}
	\label{fig:8}
	\vspace{10pt} 
\end{figure}

Next, we investigate how this regular BH responds to perturbations from a massless scalar field $ \Phi $. The scalar field's dynamics may be described using the Klein-Gordon (KG) equation:
\begin{equation}\label{eq3}
\frac{1}{\sqrt{-g}} \partial_\mu\left(\sqrt{-g} g^{\mu \nu} \partial_\nu \Phi\right) =0\,,
\end{equation}
where $g_{\mu \nu}$ is the background metric. We will analyze the eigenvalue problem associated with the dynamics of the scalar field in the frequency domain. Given the hyperspherical symmetry of the spacetime under investigation, we will use hyperspherical harmonics to separate variables for the perturbation field $\Phi$. Thus, the perturbation field $\Phi$ can be expressed as:
\begin{equation}
	\Phi(t,r,\theta_1,\theta_2,...,\theta_n)=\int\, d\omega\sum_{l,m}r^{-(d-2)/2}\Psi_{l m}(r)Y_{l m}(\theta_1,\theta_2,...,\theta_n)e^{-i \omega t}\,,
\end{equation}
where $Y_{l m}(\theta_1,\theta_2,...,\theta_n)$ is the hyperspherical harmonics, with $l$ and $m$ representing the angular and azimuthal quantum numbers, respectively. Then, the KG equation \eqref{eq3} can be transformed into a \text {Schrödinger-like} form:
\begin{equation}\label{6}
	\frac{d^2\Psi}{dr_*^2}+(\omega^2-V(r_*))\Psi(r_*)=0\,.
\end{equation}
Here, $r_*$ is the tortoise coordinate, defined by the relation $r$ by $dr_{*}/dr=1/f(r)$. The effective potential $ V $ in the above equation is given by:
	\begin{align}\label{7}
	&V(r)=  f(r)\left(\frac{l(l+d-3)}{r^2}\right. 
	 \left.+\frac{(d-2)(d-4)}{4 r^2} f(r)+\frac{d-2}{2 r} \frac{d f(r)}{d r}\right)\,, 
\end{align}
where the angular quantum number $l$ takes the integer values of $0, 1, \ldots$. For simplicity, we have removed the subscript $l$ and $m$ from $\Psi_{l,m}(t,r)$ and denoted it as $\Psi$ in Eq.\eqref{6}. In the effective potential $V$, the first and the second terms are the centrifugal potential, while the third term corresponds to the gravitational potential. In $4$-dimensional spacetime, when $l=0$, the centrifugal potential vanishes. While for $d>4$, the centrifugal potential persists even when $l=0$. Fig.\ref{fig:7} and Fig.\ref{fig:8} illustrate the effective potentials $V(r)$ for various values of the parameter $A$ and different spacetime dimension $d$. The potentials are positive definite, ensuring the stability of the scalar field perturbations.


\section{Quasinormal modes}\label{section3}

In this section, we will explore the characteristics of the QNMs associated with this regular BH, with a particular focus on the influence of the parameter $A$ and the spacetime dimension $d$. The QNMs are determined by solving an eigenvalue problem that requires a purely outgoing wave at spatial infinity ($ r_*\rightarrow+\infty $) and a purely ingoing wave at the event horizon ($ r_*\rightarrow-\infty $), specially,
\begin{eqnarray}
		\Psi \sim e^{+i\omega r_*}, & \quad r_*\rightarrow+\infty\,, 
  \label{9-1}
  \\
		\Psi \sim e^{-i\omega r_*}, & \quad r_*\rightarrow-\infty\,.
  \label{9-2}
\end{eqnarray}
The boundary conditions described above characterize the black hole’s response to a transient perturbation after the source has stopped acting \cite{Konoplya:2011qq,Berti:2009kk,Kokkotas:1999bd}.

Numerous techniques have been developed to calculate QNMs, including the WKB method \cite{Ferrari:1984zz,Schutz:1985km,Iyer:1986np,Iyer:1986nq,Konoplya:2003ii,Matyjasek:2017psv}, the asymptotic iteration  method (AIM) \cite{Ciftci:2005xn,Cho:2009cj,Cho:2011sf}, the Horowitz-Hubeny method \cite{Horowitz:1999jd}, the continued fraction method (CFM) \cite{Leaver:1985ax}, and the PSM \cite{boyd2001chebyshev,Jansen:2017oag}. In this paper, we utilize the PSM to compute the QNM spectra, as it is a highly effective numerical tool \cite{boyd2001chebyshev,Jansen:2017oag,Wu:2018vlj,Fu:2018yqx,Fu:2022cul,Gong:2023ghh,Fu:2023drp,Zhang:2024nny,Liu:2021fzr,Destounis:2021lum,Jaramillo:2021tmt,Xiong:2021cth,Song:2024kkx}, particularly adept at identifying overtone modes \cite{Fu:2023drp,Gong:2023ghh,Zhang:2024nny}.

When applying the PSM, it is advantageous to use the Eddington-Finkelstein (EF) coordinate, as this renders the wave equation \eqref{6} linear with respect to the frequency $\omega$. To achieve this, we perform the following coordinate transformation:
\begin{eqnarray}
		&&r\rightarrow(1/u^{d-3}-A)^{1/(d-3)}\,,
  \nonumber
  \\
		&&v=t+r_*\,.
  \label{coor-trans-EF}
\end{eqnarray}
Under this transformation, the metirc \eqref{metric} becomes
\begin{equation}
	ds^2_{EF}=-f(u)dv^2-2dv du(1-Au^{d-3})^{-(d-4)/(d-3)}u^{-2}+(u^{3-d}-A)^{2/(d-3)}d\Omega_{d-2}\,,\label{metric-EF}
\end{equation}
where $f(u)=1-u^d$.

Now, based on the hyperspherical harmonics, the perturbation field $\Phi$ in the EF coordinate can be expressed as follows:
\begin{equation}
	\Phi(v,u,\theta_1,\theta_2,...,\theta_n)=\int\, d\omega\sum_{l,m}u^{(d-4)/2}\psi_{l m}(u)Y_{l m}(\theta_1,\theta_2,...,\theta_n)e^{-i \omega v}\,.
 \label{sv-EF}
\end{equation}
Then, the wave equation \eqref{6} turns into the following form:
\begin{eqnarray}
		&&
		u^2 f(u) (A u^{d-3}-1)^2 \psi''(u)+ (A u^{d-3}-1) \Big[2(d-3)A u^{d-2}f(u) 
		\nonumber
		\\
		&&
		+u^2(Au^{d-3}-1)f'(u)-2i\omega(1-Au^{d-3})^{1/(d-3)}\Big] \psi'(u)+\Big[-l(d+l-3) 
		\nonumber
		\\
		&&
		-2i\omega u^{-1}(1-Au^{d-3})^{1/(d-3)}(1+(d-4)Au^{d-3}/2)+(Au^{d-3}-1) \big((d-2)(d-4)/4 
		\nonumber
		\\
		&&
		+(5+3(d-5))(d-4)Au^{d-3}/4\big)f(u)+u(d-4)(Au^{d-3}-1)f'(u)/2 \Big]	\psi(u)=0\,.
	\label{eqq15}
\end{eqnarray}
Combining with the boundary conditions \eqref{9-1} and \eqref{9-2}, we can solve Eq. \eqref{eqq15} using the PSM. 
For a detailed introduction to the PSM, please refer to Appendix \ref{app:A}.

\subsection{Fundamental modes}

\begin{figure}[htbp]
  \centering
  \includegraphics[width=0.45\textwidth]{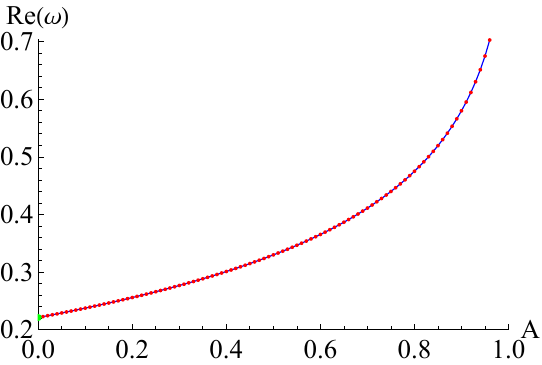}\ \hspace{0.6cm}
  \includegraphics[width=0.45\textwidth]{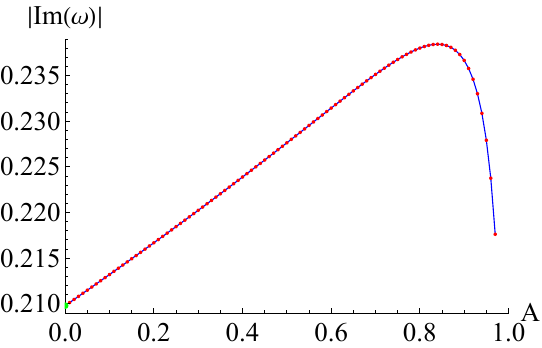}\ \\
  \caption{QNFs in $4$-dimensional spacetime as a function of the parameter $A$ for $l=0$ and $n=0$. The green point represents the Schwarzschild case, where $A=0$, while the red points correspond to various values of the parameter $A$.}
  \label{fm-d4-l0-n0}
\end{figure}
\begin{figure}[htbp]
  \centering
  \includegraphics[width=0.45\textwidth]{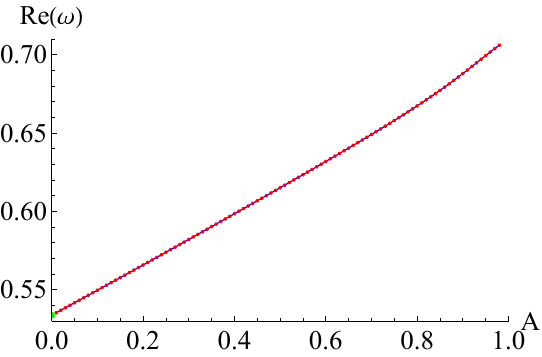}\ \hspace{0.6cm}
  \includegraphics[width=0.45\textwidth]{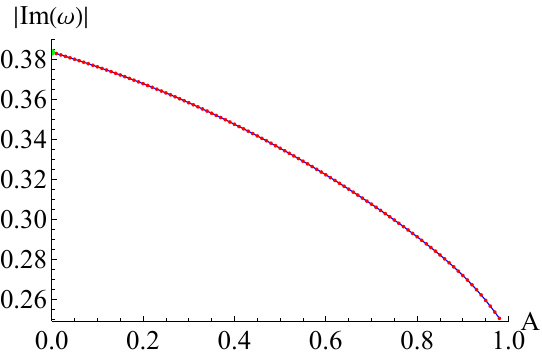}\ \\
  \caption{QNFs in $5$-dimensional spacetime as a function of the parameter $A$ for $l=0$ and $n=0$. The green point represents the Schwarzschild case, where $A=0$, while the red points correspond to various values of the parameter $A$.}
  \label{fm-d5-l0-n0}
\end{figure}

We firstly discuss the case of $l=0$. Fig.\ref{fm-d4-l0-n0} and Fig.\ref{fm-d5-l0-n0} illustrate QNFs of fundamental modes as a function of the parameter $A$ in $4$-dimensional and $5$-dimensional spacetimes, respectively. In $4$-dimensional spacetime, the real part of QNFs, $\text{Re}(\omega)$, increases as $A$ increases. Notably, as $A$ approaches $1$, $\text{Re}(\omega)$ experiences a more rapid rise. This suggests that the parameter $A$ strengths the oscillation. However, we observe that the absolute value of the imaginary part $|\text{Im}(\omega)|$  shows a non-monotonic change with respect to $A$. Specifically, as $A$ increases, $|\text{Im}(\omega)|$ first increases, signifying a faster decay modes of this system, and then rapidly decreases when $A$ approaches $1$. In contrast, in the $5$-dimensional spacetime, as $A$ increases, $\text{Re}(\omega)$ increases almost linearly. Meanwhile, $|\text{Im}(\omega)|$ decreases with increasing $A$, which differs from the behavior observed in the $4$-dimensional case. 
\begin{figure}[htbp]
  \centering
  \includegraphics[width=0.45\textwidth]{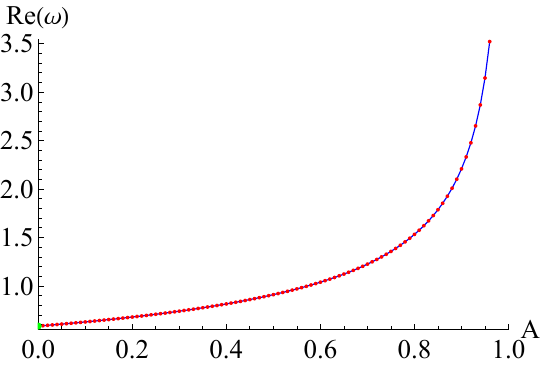}\ \hspace{0.6cm}
  \includegraphics[width=0.45\textwidth]{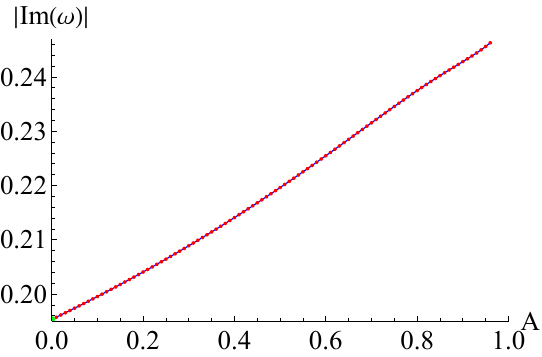}\ \\
  \caption{QNFs in $4$-dimensional spacetime as a function of the parameter $A$ for $l=1$ and $n=0$. The green point represents the Schwarzschild case, where $A=0$, while the red points correspond to various values of the parameter $A$.}
  \label{fm-d4-l1-n0}
\end{figure}
\begin{figure}[htbp]
  \centering
  \includegraphics[width=0.45\textwidth]{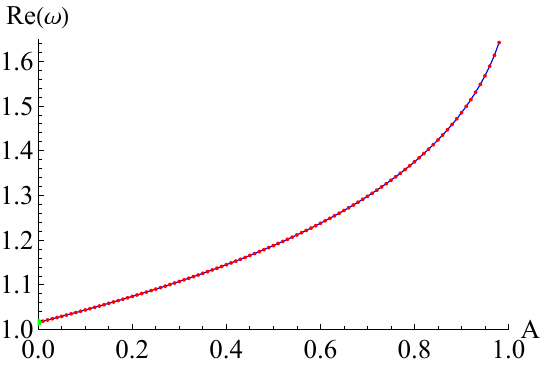}\ \hspace{0.6cm}
  \includegraphics[width=0.45\textwidth]{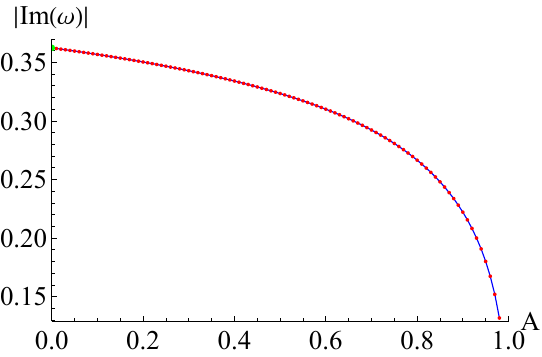}\ \\
  \caption{QNFs in $5$-dimensional spacetime as a function of the parameter $A$ for $l=1$ and $n=0$. The green point represents the Schwarzschild case, where $A=0$, while the red points correspond to various values of the parameter $A$.}
  \label{fm-d5-l1-n0}
\end{figure}

When the angular quantum number $l$ is turned on, it is observed that as $A$ approaches $1$, the real part $\text{Re}(\omega)$ of QNFs in $4$-dimensional spacetime exhibits a more rapid increase compared to the $l=0$ case (left plot in Fig.\ref{fm-d4-l1-n0}). In the $5$-dimensional spacetime, in sharp contrast to the $l=0$ scenario, $\text{Re}\omega$ also rapidly increase as $A$ approaches $1$ when $l$ is turned on (left plot in Fig.\ref{fm-d5-l1-n0}). Additionally, the previously non-monotonic behavior of $\text{Im}\omega$ in $4$-dimensional spacetime for $l=0$ becomes almost linear increasing once $l$ is turned on (right plot in Fig.\ref{fm-d4-l1-n0}), while in $5$-dimensional spacetime, the decrease becomes more rapid as $A$ approaches $1$ compared to the $l=0$ case (right plot in Fig.\ref{fm-d5-l1-n0}).

Typically, the fundamental modes are associated with variations in the peak of the effective potential, allowing us to understand the behaviors of QNFs in different scenarios. In $4$-dimensional spacetime, the peak values and their positions change significantly as $A$ increases, particularly as $A$ approaches $1$ for $l=1$ compared to $l=0$. However, this relative change between $l=0$ and $l=1$ in $5$-dimensional spacetime is less pronounced. These changes in the peak of the effective potential are consistent with the variations in the QNFs of the fundamental modes.

\begin{figure}[htbp]
  \centering
  \includegraphics[width=0.55\textwidth]{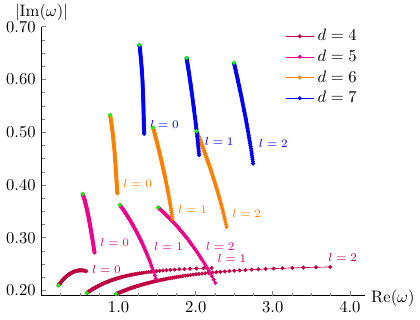}\ \\
  \caption{The phase diagram $ \text{Re}(\omega)-|\text{Im}(\omega)| $ of the fundamental modes for various spacetime dimensions $d$ and angular quantum numbers $l$. Green points represent the Schwarzschild case, i.e., $A=0$, while the finial points correspond to $A=0.9$.}
  \label{fig:2}
\end{figure}

To further clearly illustrate the effects of spacetime dimension $d$ and the angular quantum number $l$, we also show the phase diagram of the real part $ \text{Re}(\omega) $ and the absolute value of the imaginary part $ |\text{Im}(\omega)| $ of the fundamental modes for various $d$ and $l$ in Fig.\ref{fig:2}. Green points represent the Schwarzschild case, i.e., $A=0$, while the finial points correspond to $A=0.9$. The non-monotonic behavior is visible only in $\text{Im}\omega$ for $l=0$ and $d=4$. For $l>0$ and $d>4$, this non-monotonic behavior disappears, likely due to the presence of a centrifugal potential that significantly alters the shape of the effective potential near its peak, suppressing the effect of the parameter $A$. Moreover, when both $l$ and $d$ are fixed, the real oscillation frequency of the fundamental mode increases monotonically with $A$. This increase is more pronounced in $4$-dimensional spacetime compared to higher dimensions, suggesting that the influence of $A$ on oscillatory behaviors reduces as $d$ increases. Additionally, as $A$ increases, the absolute value of the imaginary part $ |\text{Im}(\omega)| $ rapidly decreases for dimensions beyond $d=4$, indicating that the impact of $A$ on the decay behavior of the QNFs becomes significant in higher dimensions.

\subsection{Overtones}
\begin{figure}[htbp]
  \centering
  \includegraphics[width=0.45\textwidth]{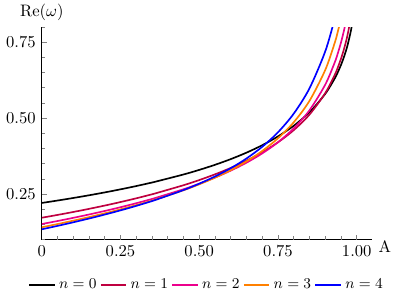}\ \hspace{0.6cm}
  \includegraphics[width=0.45\textwidth]{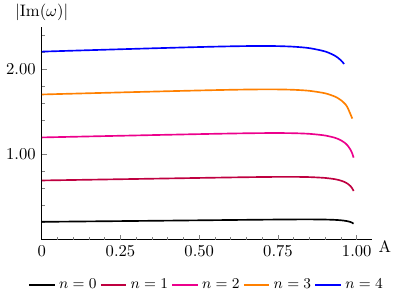}\ \\
  \caption{QNFs of the overtones in $4$-dimensional spacetime as a function of the parameter $A$ for $l=0$.}
  \label{om-d4-l0-n}
\end{figure}

In this subsection, we explore the characteristics of the overtones. Fig.\ref{om-d4-l0-n} displays the QNFs of the overtones in $4$-dimensional spacetime as a function of the parameter $A$ for $l=0$. It is evident that no overtone outburst is observed in this model with $d=4$. Instead, the QNFs of the overtones exhibit similar behaviors to the fundamental modes with respect to $A$. What’s slightly different is that as $A$ approaches $1$, $\text{Re}(\omega)$ increases rapidly while $|\text{Im}(\omega)|$ decreases with increasing overtone number $n$. 

\begin{figure}[htbp]
  \centering
  \includegraphics[width=0.45\textwidth]{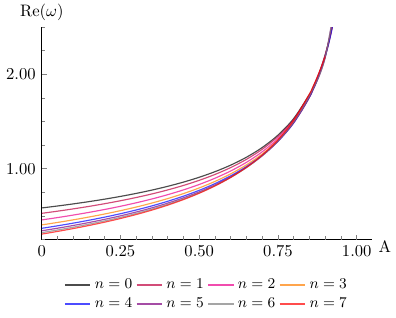}\ \hspace{0.6cm}
  \includegraphics[width=0.45\textwidth]{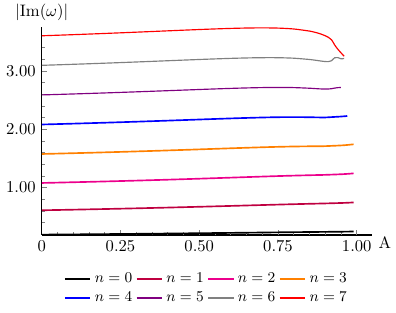}\ \\
  \caption{QNFs of the overtones in $4$-dimensional spacetime as a function of the parameter $A$ for $l=1$.}
  \label{om-d4-l1-n}
\end{figure}

Next, we turn to study the case of $l=1$ (see Fig.\ref{om-d4-l1-n}). We find that the behaviors of $\text{Re}(\omega)$ for the overtones is also similar to that of $l=0$. However, compared to the $l=0$ case, the non-monotonic behavior in $|\text{Im}(\omega)|$ vanishes for overtone numbers $n=1,2,3,4,5$. This non-monotonic behavior reemerges for $n=6,7$. This observation suggests that the angular quantum number $l$ suppresses the effects of quantum gravity or modified gravity. Similar observations have been reported in previous studies \cite{Gong:2023ghh,Zhang:2024nny,Song:2024kkx}.

\begin{figure}[htbp]
  \centering
  \includegraphics[width=0.45\textwidth]{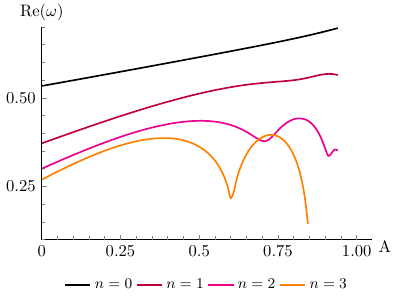}\ \hspace{0.6cm}
  \includegraphics[width=0.45\textwidth]{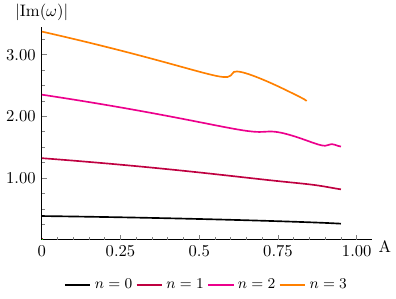}\ \\
  \caption{QNFs of the overtones in $5$-dimensional spacetime as a function of the parameter $A$ for $l=0$.}
  \label{om-d5-l0-n0}
\end{figure}
\begin{figure}[htbp]
  \centering
  \includegraphics[width=0.45\textwidth]{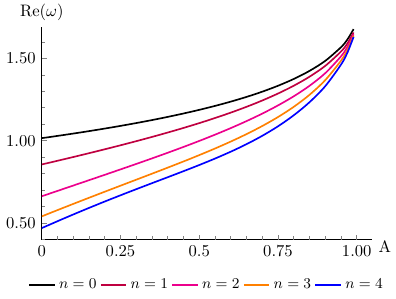}\ \hspace{0.6cm}
  \includegraphics[width=0.45\textwidth]{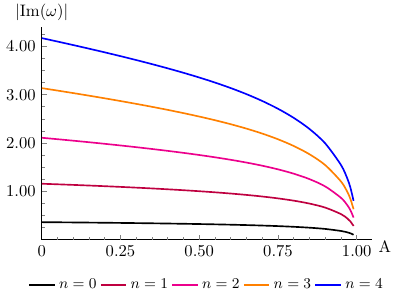}\ \\
  \caption{QNFs of the overtones in $5$-dimensional spacetime as a function of the parameter $A$ for $l=1$.}
  \label{om-d5-l1-n0}
\end{figure}

Now, we turn our attention to investigating the properties of overtones in $5$-dimensional spacetime. As depicted in Fig.\ref{om-d5-l0-n0}, the QNFs of overtones in $5$-dimensional spacetime are shown as a function of the parameter $A$ for $l=0$. In contrast to the case in $4$-dimensional spacetime, an obvious overtone outburst is observed in $5$-dimensional spacetime for higher overtones ($n=2,n=3$). The quantum gravity effect is the cause behind this overtone outburst. This phenomenon has been widely noticed in modified gravity theories and effective quantum gravity models such as \cite{Berti:2005ys,Berti:2018vdi,Fu:2022cul,Fu:2023drp,Gong:2023ghh,Moura:2021eln,Moura:2021nuh,Moura:2022gqm,Lin:2024ubg,Ghosh:2022gka,Zhang:2024nny,Konoplya:2022pbc}. After the initial outburst of overtones, an oscillatory behavior is detected. This particular pattern has been witnessed in RN-BH \cite{Berti:2003zu,Jing:2008an} and other effective quantum gravity corrected BH \cite{Fu:2023drp,Gong:2023ghh,Zhang:2024nny}. However, when the angular quantum number $l$ is activated, that is, $l > 0$, no overtone outburst is detected as shown in Fig.\ref{om-d5-l1-n0}. This further demonstrates that the angular quantum number $l$ has a suppressive effect on the quantum gravity effect. Finally, for the convenience of readers, we also provide the QNF values for various spacetime dimensions $d$ as well as for different values of $A$, $l$ and $n$ in TABLE \ref{table3}.

\begin{table}[htbp]
	\centering
	\fontsize{11}{11}\selectfont
	\begin{tabular}{|c|c|cc|cc|} 
		\hline
		& & \multicolumn{2}{c|}{$ l=0 $}  & \multicolumn{2}{c|}{$ l=1 $}\\ 
		\hline
		$ d $ &$ n $& $ A=0 $  & $ A=0.2 $ & $ A=0 $ & $ A=0.2 $ \\ 
		\hline
		4 & 0 & 0.22091-0.20979i & 0.25562-0.21668i & 0.58587-0.19532i & 0.67905-0.20406i \\
		& 1 & 0.17222-0.69611i & 0.21233-0.70796i & 0.52890-0.61251i & 0.63092-0.63392i  \\
		& 2 & 0.15156-1.20264i & 0.19454-1.21803i & 0.45908-1.08027i & 0.56983-1.10604i\\
		& 3 & 0.13474-1.70522i & 0.18667-1.72678i & 0.40650-1.57660i & 0.52200-1.60459i\\
		
		5 & 0 & 0.53384-0.38338i & 0.56584-0.36764i & 1.01602-0.36233i & 1.07415-0.35503i \\ 
		& 1 & 0.37186-1.32261i & 0.43402-1.23939i & 0.85638-1.15761i & 0.94842-1.10679i\\
		& 2 & 0.30033-2.35330i & 0.37424-2.19001i & 0.66344-2.10918i & 0.79353-1.98378i\\
		& 3 & 0.26939-3.37723i & 0.35095-3.13438i & 0.54135-3.13670i & 0.69099-2.92551i\\
		
		6 & 0 & 0.88944-0.53310i & 0.91790-0.50459i & 1.44651-0.50927i & 1.48985-0.48545i \\
		& 1 & 0.55035-1.89670i & 0.63988-1.74503i & 1.14354-1.64146i & 1.24912-1.54847i\\
		& 2 & 0.41458-3.49012i & 0.51145-3.18709i & 0.76082-3.11901i & 0.93661-2.87161i\\
		& 3 & 0.36933-5.04618i & 0.46602-4.60027i & 0.58272-4.73201i & 0.77050-4.31972i\\
		
		7 & 0 & 1.27054-0.66578i & 1.29536-0.62892i & 1.88140-0.64108i & 1.91560-0.60953i \\
		& 1 & 0.68343-2.43879i & 0.80902-2.22137i & 1.39266-2.06571i & 1.52289-1.94852i\\
		& 2 & 0.50421-4.63695i & 0.61231-4.20853i & 0.76366-4.17660i & 0.98524-3.79518i\\
		& 3 & 0.45619-6.72636i & 0.54728-6.10319i & 0.59931-6.39379i & 0.79794-5.79251i\\
		
		8 & 0 & 1.66869-0.78557i & 1.69018-0.74283i & 2.32073-0.76102i & 2.34835-0.72396i \\
		& 1 & 0.75402-2.98472i & 0.91847-2.68216i & 1.60031-2.42111i & 1.76649-2.29865i\\
		& 2 & 0.58573-5.79782i & 0.69558-5.25365i & 0.74822-5.32442i & 0.98259-4.80134i\\
		& 3 & 0.54003-8.41535i & 0.61111-7.63003i & 0.63532-8.09335i & 0.82107-7.32006i\\
		\hline
	\end{tabular}
	\centering
	\caption{The QNMs for the massless scalar field perturbation with different values of $ d $, $ n $, $ l $ and $ A $ by PSM.}\label{table3}
\end{table}

\section{Conclusions and discussions}\label{section4}

In this paper, we systematically investigated the properties of the QNMs associated with a probe scalar field over a regular BH featuring an integrable singularity, which is characterized by the parameter $A$. Our primary focus is on how the QNFs vary with changes in the parameter $A$. We also explore the influence of the angular quantum number $l$ and the spacetime dimension $d$ on the QNFs. 

In $4$-dimensional spacetime, a non-monotonic behavior in the absolute value of the imaginary part of the fundamental mode emerges with respect to the parameter $A$ for vanishing angular quantum number $l$. This behavior can be ascribed to the effective quantum gravity effects, as discussed in references \cite{Fu:2022cul,Fu:2023drp,Gong:2023ghh,Zhang:2024nny,Song:2024kkx}. Notably, when the angular quantum number $l$ is turned on, this non-monotonic behavior ceases to exist, suggesting that the influence of the angular quantum number $l$ supersedes the effects of effective quantum gravity \cite{Fu:2022cul,Fu:2023drp,Gong:2023ghh,Zhang:2024nny,Song:2024kkx}. Furthermore, our analysis reveals that this non-monotonic behavior reappears in higher overtones, indicating that the effective quantum gravity effects regain dominance in these higher overtones. 

An additional significant finding of this study is the observation that an overtone outburst accompanied by an oscillatory pattern occurs exclusively when the dimensionality $d$ exceeds $4$. This phenomenon is notably absent in the scenario where $d = 4$. This distinction underscores the critical role of dimensionality in influencing the dynamic behavior of the system under investigation. The underlying reasons for this distinction warrant further investigation in future research.

\acknowledgments

This work is supported by National Key R$\&$D Program of China (No. 2020YFC2201400), the Natural Science Foundation of China under Grants No. 12375055, 12447151 and 12347159.

\appendix
\section{Pseudo-spectral method}\label{app:A}

In this appendix, we provide a detailed introduction on the PSM used in Section \ref{section3}. For a more comprehensive introduction, readers may refer to \cite{Boyd:Chebyshev,Jansen:2017oag} and additional references \cite{Fu:2022cul,Gong:2023ghh,Zhang:2024nny}.

In EF coordinate system, Eq. \eqref{eqq15} automatically fulfills the ingoing boundary condition \eqref{9-2} at the horizon. Therefore, we only need to impose the outgoing boundary condition \eqref{9-1} at spatial infinity. To achieve this, we implement the following transformation:
\begin{eqnarray}
    &\psi(u) = e^{2i\omega/u}u^{-2i\omega}\delta\psi(u)\,,\,\, &d = 4\,,\\
    &\psi(u)= e^{2i\omega/u}\delta\psi(u)\,,\qquad\,\,\, &d > 4\,.
\end{eqnarray}
which ensure that Eq. \eqref{eqq15} satisfies the outgoing condition at the spatial infinity. Under the above transformation, Eq. \eqref{eqq15} can be rewritten as:
\begin{eqnarray}
\label{reweq}
    \delta\psi''(u)+\lambda_0(u,\omega)\delta\psi'(u)+s_0(u,\omega)\delta\psi(u)=0\,.
\end{eqnarray}
The coefficients $\lambda_0(u,\omega)$ and $s_0(u,\omega)$ for $d=4$ are:
\begin{eqnarray}
        \lambda_{0(d=4)}&=&\frac{f'(u)}{f(u)}+\frac{2A}{Au-1}-\frac{4i\omega}{u^2}-\frac{4i\omega}{u}+\frac{2i\omega}{u^2f(u)}\,, 
        \label{lambda-de4}
        \
        \\
        s_{0(d=4)}&=&-\frac{l(l+1)}{u^2(Au-1)^2f(u)}-\frac{4(1+u)\omega((1+u)f(u)-1)}{u^4f(u)}
        \nonumber
        \\
        &&
        -\frac{(1+u)f'(u)}{u^2f(u)}-\frac{2i\omega((Au^2+u+2)f(u)-1)}{u^3(Au-1)f(u)}\,,
        \label{s-de4}
\end{eqnarray}
and for $d>4$ they are:
\begin{eqnarray}
 \lambda_{0(d>4)}&=&\frac{f'(u)}{f(u)}+\frac{2A(d-3)}{Au-u^{4-d}}-\frac{4i\omega}{u^2}+\frac{2i\omega u(1-Au^{d-3})^{\frac{1}{d-3}}}{(u^3-Au^d)f(u)}\,,
 \label{lambda-dl4}
    \
    \\
s_{0(d>4)}&=&\frac{(d-4)f'(u)}{2uf(u)}-\frac{l(l+d-3)u^4}{(u^3-Au^d)^2f(u)}-\frac{(d-4)(u^3(d-2)+Au^d(3d-10))}{4u^2(u^3-Au^d)}
\nonumber
\\
&&
-i\omega\left[-\frac{2f'(u)}{u^2f(u)}+\frac{4(u^3+Au^d(d-4))}{u^6-Au^{3+d}}-\frac{(1-Au^{d-3})^{\frac{1}{d-3}}(2u^3+Au^d(d-4))}{(u^3-Au^d)^2f(u)}\right]
\nonumber
\\
&&
+4\omega^2\left[\frac{u^{\frac{d}{3}}(u^3-Au^d)^{1-\frac{1}{d-3}}}{u^2f(u)}-\frac{1}{u^4}\right]\,.
\label{s-dl4}
\end{eqnarray}
It is evident that the rewritten wave equation \eqref{reweq} satisfies the ingoing boundary condition at the horizon and the outgoing boundary condition at spatial infinity.

The core aspect of the PSM is to discretize the linear differential equation \eqref{reweq} and subsequently
solve the resulting generalized eigenvalue problem. The first step in this process is to replace the continuous variables with a discrete set of collocation points, known as the grid points. By doing so, a function can be represented by the values it takes at these grid points. Specially, the function $\delta\psi(u)$ in Eq.\eqref{reweq} can be expanded into the following form: 
\begin{eqnarray}
    \delta\psi(u)=\displaystyle\sum^N_{j=0}\alpha(u_j)C_j(u)\,,
\end{eqnarray}
where $u_j$ are the grid points, and the polynomials $C_j(u)$ are the cardinal functions associated with these grid points, satisfying the condition $C_j(u_i)=\delta_{ij}$. Here, $i,j=0,\ldots\, N$. The choice of a specific grid uniquely determines the cardinal functions $C_j(u)$. They are given by:
\begin{eqnarray}
\label{Cju}
		C_j(u)=\prod_{j=0,j\neq i}^N \frac{u-u_j}{u_i-u_j}\,.
\end{eqnarray}
By employing cardinal functions, we can contrast the matrix $D_{ij}^{(1)}$ that represents the first derivative, defined as $D_{ij}^{(1)}=C'_i(u_j)$. Similarly, we can further construct the matrices for higher derivatives.

The successful implementation of this method hinges critically on the choice of collocation points. Among various options, the Chebyshev grid is commonly preferred and frequently utilized due to its superior performance. The Chebyshev grid is defined as:
\begin{eqnarray}
\label{ui-cos}
    u_i=\cos\left(\frac{i}{N}\pi\right)\,,
\end{eqnarray}
For this grid, the cardinal functions $C_j(u)$ can be expressed as linear combinations of Chebyshev polynomials $T_n(x)$:
\begin{eqnarray}
    C_j(u)=\frac{2}{N p_j}\displaystyle\sum^N_{m=0}\frac{1}{p_m}T_m(u_j)T_m(u)\,,
\end{eqnarray}
where $p_j$ are the weighting factors. Specifically, $p_j=1$ for $j\neq 0$ and $j\neq N$, while $p_0=p_N=2$. These weighting factors play a crucial role in ensuring numerical stability.
The cardinal functions $C_j(u)$ play a central role in the discretization process, as they provide a smooth and accurate basis for the approximation of functions on the Chebyshev grid.

Usually, the coefficients $\lambda_0(u,\omega)$ and $s_0(u,\omega)$ are the polynomials in the frequency\footnote{In the present model, for $d=4$, the coefficients $\lambda_0(u,\omega)$ and $s_0(u,\omega)$ are linear in $\omega$. For $d>4$, $\lambda_0(u,\omega)$ remains linear in $\omega$, whereas $s_0(u,\omega)$ includes a dependence on $\omega^2$.}. These polynomials can be expressed as the form $\displaystyle\sum_{p}c_p(u)\omega^p$. To facilitate numerical computations, the coefficients $c_p(u)$ are discretized at the grid points, thereby converting them into vectors. These vectors are then multiplied by the corresponding derivative matrices $D_{ij}^{(n)}$ and the resulting matrices are summed. This process enables the original equation \eqref{reweq} to be reformulated as the following generalized eigenvalue equation:
\begin{eqnarray}
\label{app-2}
    \mathcal{M}(\omega)\delta\psi=0\,,
\end{eqnarray}
where $\mathcal{M}(\omega)\equiv M_0+\omega M_1+\omega^2M_2+\cdots+\omega^pM_p$ and $M_i$ denotes the linear combination of the derivative matrices. Notice that $M_i$ are purely numerical matrices.
Through these transformations, the linear ordinary differential equation \eqref{reweq}, subject to specific boundary conditions, is converted into a matrix equation.
By numerically solving the above eigenvalue equation \eqref{app-2}, we can determine the QNFs.

\bibliographystyle{style1}
\bibliography{qnm}

\end{document}